\documentclass{article}
\usepackage{spconf,amsmath,graphicx,hyperref}
\usepackage{booktabs}
\usepackage{multirow}
\usepackage[numbers]{natbib}
\title{SIGNED GRAPH UNLEARNING}
\name{Zhifei Luo$^{\star}$ \qquad Lin Li$^{\star}$ \qquad Xiaohui Tao$^{\dagger}$ \qquad Kaize Shi$^{\dagger}$}

\address{$^{\star}$ Wuhan University of Technology \\
         $^{\dagger}$ University of Southern Queensland \\
         luozhifei@whut.edu.cn, cathylilin@whut.edu.cn, Xiaohui.Tao@unisq.edu.au, Kaize.Shi@unisq.edu.au}

\begin{document}
\ninept
\maketitle
\begin{abstract}
The proliferation of signed networks in contemporary social media platforms necessitates robust privacy-preserving mechanisms. Graph unlearning, which aims to eliminate the influence of specific data points from trained models without full retraining, becomes particularly critical in these scenarios where user interactions are sensitive and dynamic. Existing graph unlearning methodologies are exclusively designed for unsigned networks and fail to account for the unique structural properties of signed graphs. Their naive application to signed networks neglects edge sign information, leading to structural imbalance across subgraphs and consequently degrading both model performance and unlearning efficiency. This paper proposes SGU (\underline{S}igned \underline{G}raph \underline{U}nlearning), a graph unlearning framework specifically for signed networks. SGU incorporates a new graph unlearning partition paradigm and a novel signed network partition algorithm that preserve edge sign information during partitioning and ensure structural balance across partitions. Compared with baselines, SGU achieves state-of-the-art results in both model performance and unlearning efficiency.
\end{abstract}
\section{Introduction}

Growing public concern regarding data privacy and the tightening of regulatory has underscored the necessity for machine learning systems to support machine unlearning-the capability to expunge the influence of particular data points from trained models. Partition-based unlearning approaches, exemplified by SISA~\cite{bourtoule2021machine}, have emerged as practical solutions in this context. These methods provide robust generalization performance while enabling efficient processing of unlearning requests dealing with image, text and so on.

Nevertheless, an increasing proportion of real-world data is naturally represented as graphs, where the relationships between entities are as crucial as the entities themselves. In particular, signed networks have attracted significant attention across diverse domains, including social media (e.g., like/dislike relationships), e-commerce (trust/distrust), and political networks (ally/enemy), where edges encode polarity information-either positive or negative-to capture the nature of interactions between nodes. To effectively model these complex relational semantics, Signed Graph Neural Networks~\cite{derr2018signed, li2020learning, huang2021sdgnn} (SGNNs) have been proposed, leveraging both the structural properties and the sign attributes of edges to learn expressive node representations.

While recent advances in graph unlearning-exemplified by partition-based methods like GraphEraser~\cite{chen2022graph} and non-SISA approaches~\cite{wu2023gif,chien2022certified,wu2023certified,cheng2023gnndelete,li2024towards}-have demonstrated efficiency in removing training data from GNN models, these techniques are primarily designed for unsigned graphs. Crucially, existing methods neglect critical edge polarity considerations during partitioning. When directly applied to signed networks, they incur substantial information loss (e.g., sign semantics) and uneven edge distributions across partitions. These limitations lead to significant performance degradation in downstream tasks and compromised unlearning efficiency, highlighting the need for specialized frameworks tailored to signed graph structures.

To address the aforementioned challenges, we introduce \textit{SGU} (\underline{S}igned \underline{G}raph \underline{U}nlearning), a SISA-style framework specifically designed for signed graph neural networks (SGNNs). Unlike traditional methods, SGU takes edge polarity into account during the partitioning process. The core idea of SGU is to identify cohesive node groups connected by positive edges while preserving external negative edges, and then use agglomerative hierarchical clustering, which integrates structural proximity and edge polarity, to form balanced partitions. This approach avoids the critical loss of sign information, prevents imbalances in edge distributions, and mitigates significant degradation in both model performance and unlearning efficiency that would typically occur when unlearning is applied to signed networks using naive partitioning methods.


In summary, this paper makes the following contributions: 
(1) We propose SGU, the first graph unlearning framework tailored for signed graphs. Concretely, we introduce a bottom-up signed graph-reconstruction paradigm that performs partitioning to preserve polarity and balance. 
(2) A sign-aware balanced clustering is designed to leverage both structural proximity and edge polarity to produce semantically meaningful and well-balanced partitions. 
(3) Extensive experiments are conducted on four real-world signed graph datasets and three SGNN backbones. The results show that our SGU achieves higher model utility than traditional graph unlearning and demonstrates reliable unlearning capacity.

\begin{figure}[!t]
  \centering
  \includegraphics[width=0.8\linewidth]{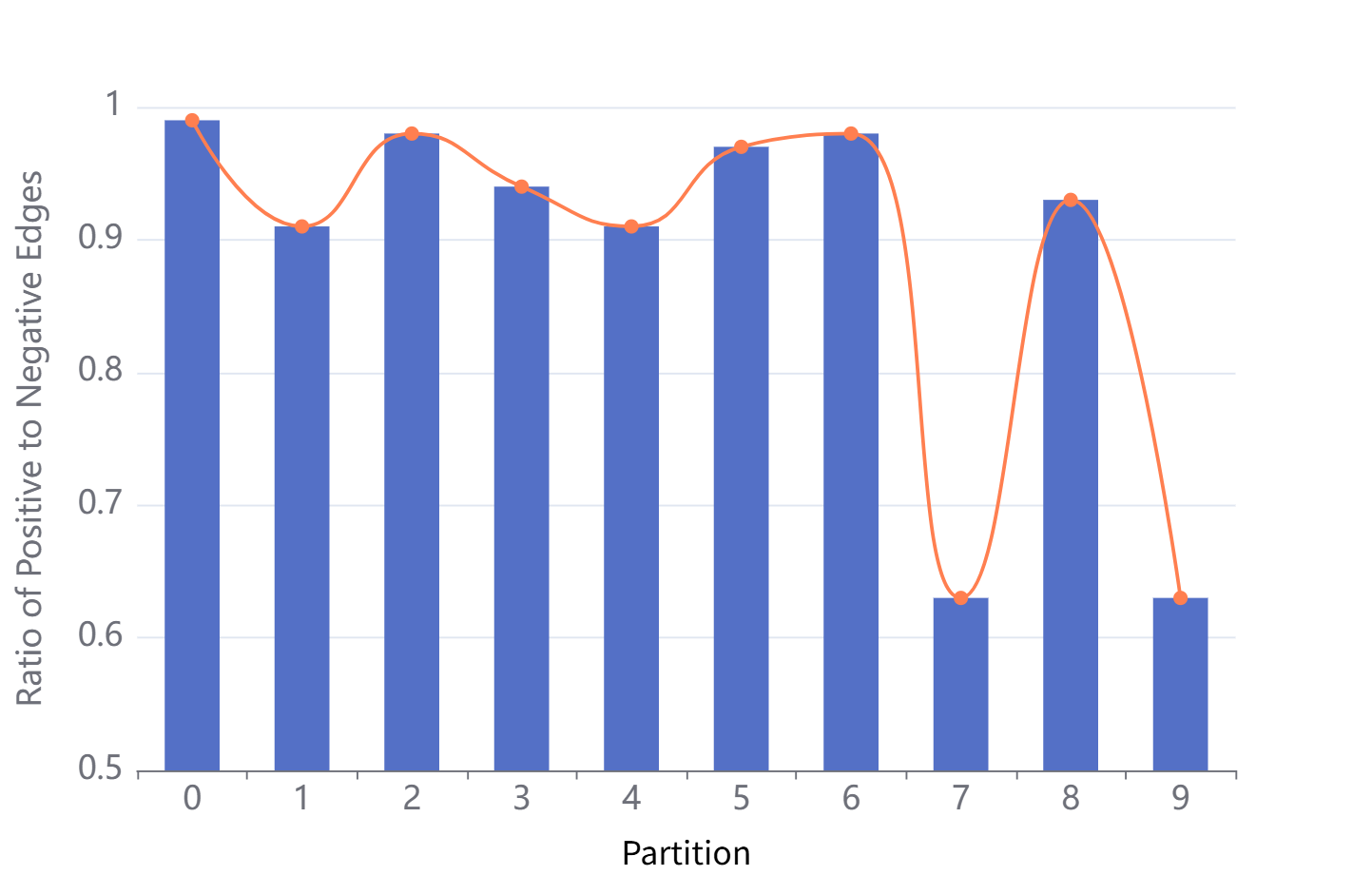}\\[4pt]
  \includegraphics[width=0.8\linewidth]{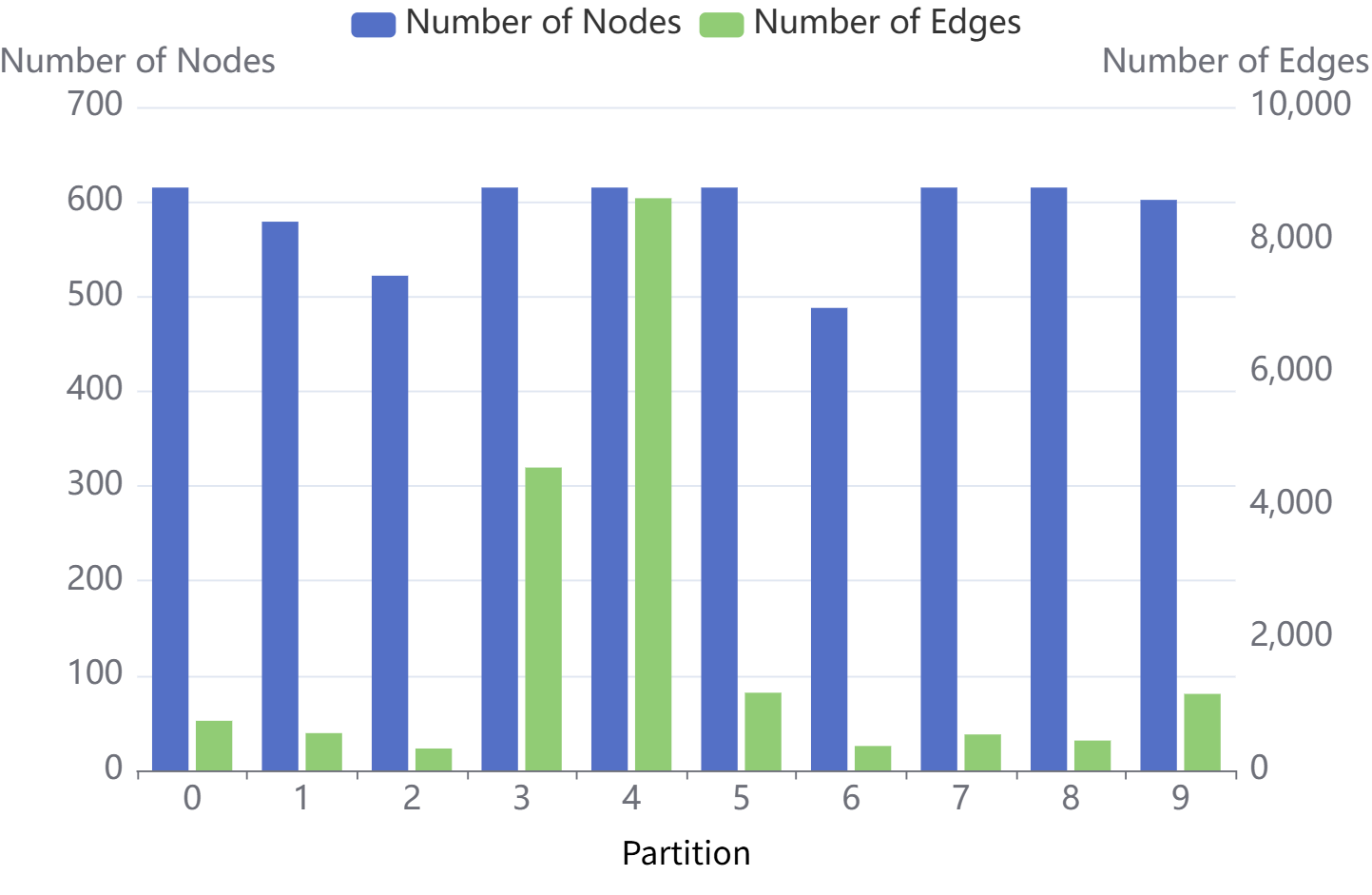}
  \caption{Partitioning results on the Bitcoin-OTC dataset using GraphEraser. The top figure shows the ratio of positive and negative edges within each partition, while the bottom figure presents the number of nodes and edges in each partition.}
  \label{fig:partitioning-results}
\end{figure}

\section{Signed Graph Unlearning}

\subsection{Signed Graph}

A signed graph is a graph where edges are assigned a positive or negative sign, representing relationships such as trust/distrust or agreement/disagreement between nodes. Formally, we define a signed graph as $G = \langle V, A^+, A^-\rangle$, where $V$ is the set of nodes, $A^+ \in \{0,1\}^{n \times n}$ and $A^- \in \{0,1\}^{n \times n}$ are the adjacency matrices for positive and negative edges, respectively.

Signed Graph Neural Networks (SGNNs) are extensions of traditional GNNs tailored for signed networks, where most existing methods are fundamentally motivated by balance theory~\cite{heider1946attitudes, cartwright1956structural}. Balance theory, originating from social psychology, suggests that signed relationships in a network tend to organize themselves in a way that minimizes social instability.
Figure 1 shows the four types of signed triangles categorized according to balance theory. In this work, we take balance theory as a starting point to explore the problem of unlearning in signed networks, aiming to understand how signed structure impacts information removal.

\begin{figure}[t]
\centering
\includegraphics[width=0.8\columnwidth]{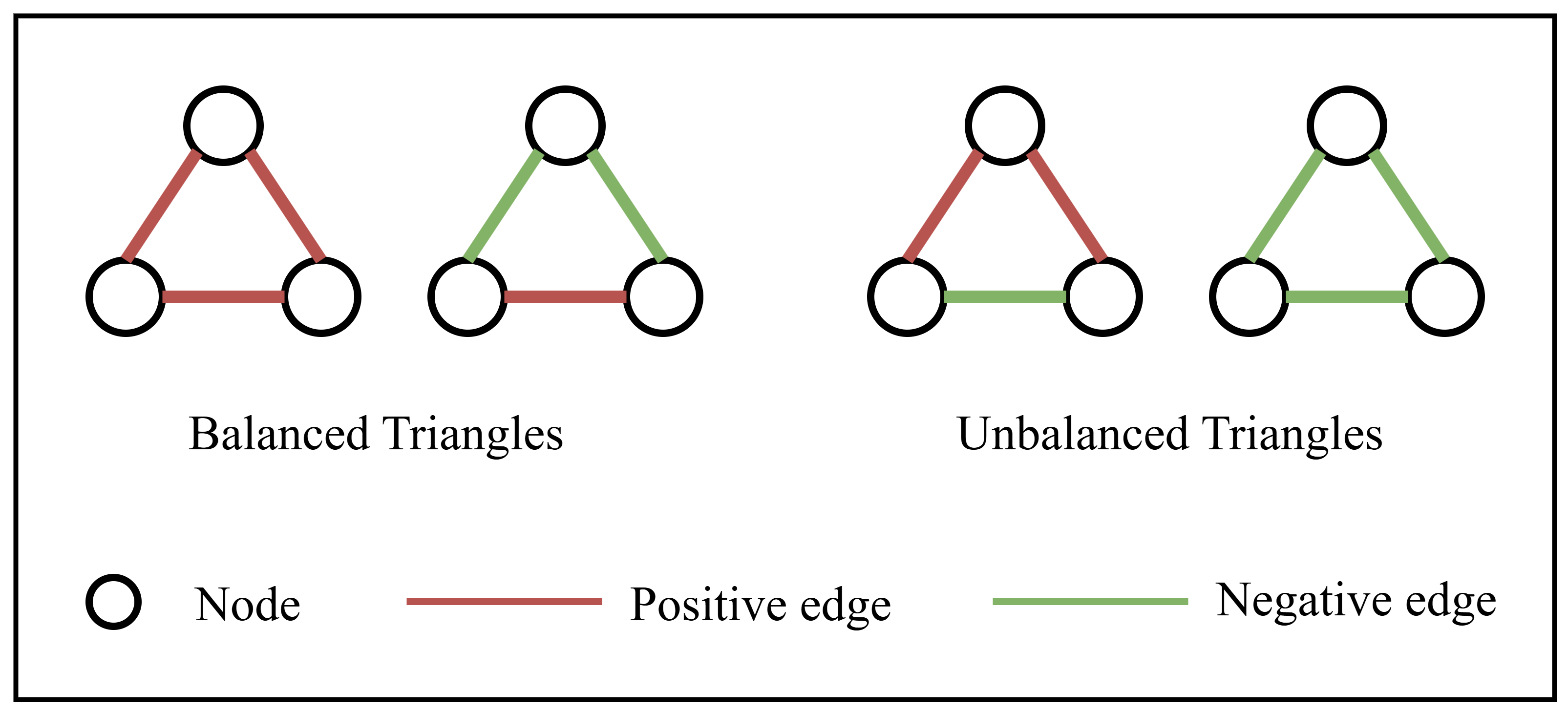}
\caption{The four undirected signed triangle types according to balance theory. The two on the left, with an even number of negative edges, are balanced; the two on the right, with an odd number, are unbalanced.}
\label{fig:triangle_types}
\end{figure}

In this paper, we focus on the classical link sign prediction problem, which is commonly used as the benchmark for signed network embedding algorithms.
This task aligns with the characteristics of most signed network datasets, which typically lack ground-truth node labels.

\subsection{Challenges of Unlearning in SGNNs}
We adopt partition-based graph unlearning methods due to their strong generalization and robustness~\cite{fan2025opengu}. However, existing partition-based methods are not directly suitable for signed graphs.

\noindent\textbf{Challenge 1: Sign Information Loss} 
Traditional partition-based methods are designed for unsigned graphs without considering edge polarity, whereas SGNNs heavily rely on edge polarity to model trust, distrust, agreement, or disagreement.

\noindent\textbf{Challenge 2: Edge Imbalance}
First, ignoring edge polarity during partitioning can lead to a severe imbalance in the ratio of positive to negative edges across different subgraphs.
Second, due to the social nature of most signed networks, applying existing methods
often results in a few subgraphs containing the majority of the original graph's edges.
Figure 2 illustrates the 
issues on the Bitcoin-OTC~\cite{kumar2016edge, kumar2018rev2} dataset.

\subsection{SGU Framework}

To address the aforementioned challenges of signed graph unlearning, we propose SGU, which consists of the following three phases: signed structure extraction, signed-aware balanced clustering, and shard model training and aggregation. The overview of SGU is illustrated in Figure 3.

\begin{figure*}[!t]
  \centering
  \includegraphics[width=\linewidth, keepaspectratio]{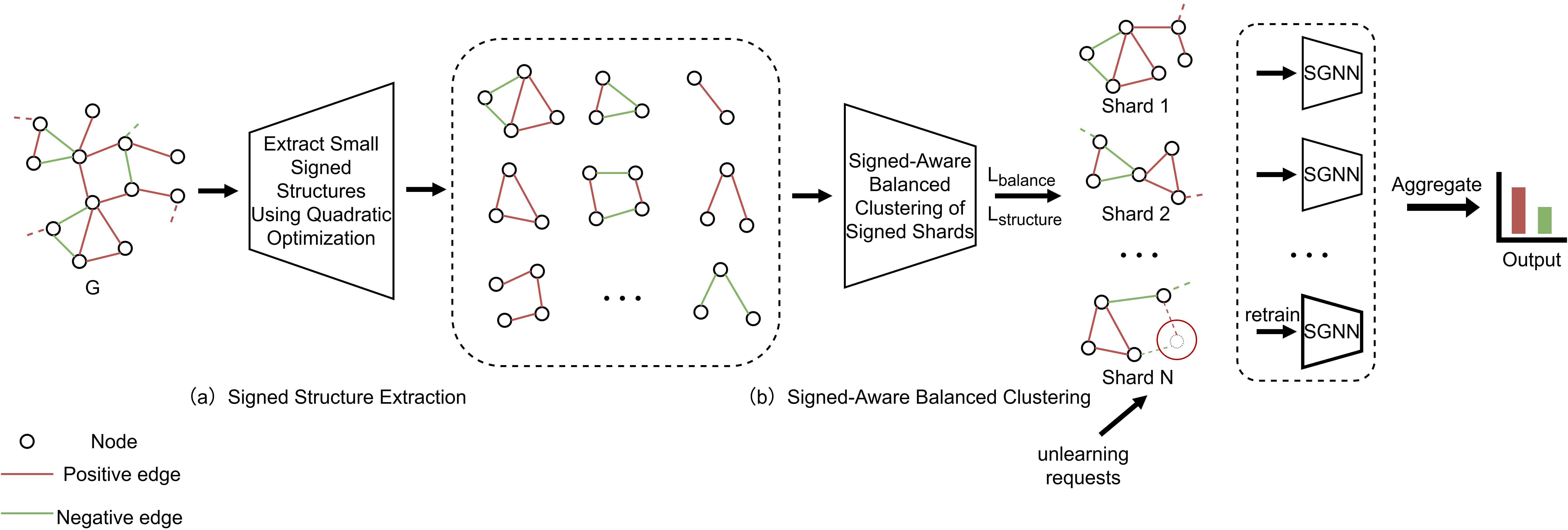}
  \caption{The Overall Framework of SGU. The Signed Structure Extraction module uses a quadratic optimization to identify balanced signed structures, ensuring the preservation of sign information and structural coherence. The Signed-Aware Balanced Clustering module generates edge-balanced and structurally complete partitions which serve as disjoint shards for training.}
  \label{fig:sgu_framework}
\end{figure*}




\section{Balanced Signed Graph Partition}

\subsection{Signed Structure Extraction}
In signed graph partitioning, there are two main lines of approaches. The first class focuses on extracting (nearly) balanced subgraphs by retaining most positive links within subgraphs and placing most negative links between subgraphs. The second class of approaches partitions the graph by identifying multiple communities. 
These communities correspond to the balanced structures grounded in balance theory discussed above, thereby preserving the full spectrum of edge polarity.


Therefore, we choose to follow the second line of approaches. Specifically, we adopt the FOCG~\cite{chu2016finding} (Finding Oppositive Cohesive Groups) algorithm. FOCG is an efficient algorithm for identifying multiple oppositive cohesive groups (k-OCGs) in signed networks.
It extracts groups, each consisting of multiple subgraphs, where there is dense cohesion within each subgraph and strong opposition between subgraphs.

\subsection{Signed-Aware Balanced Clustering}
Although FOCG effectively extracts high-quality signed structures, the resulting groups are typically small in scale and insufficient to capture the global characteristics of the original signed graph. To address this, we further cluster these groups to obtain a suitable number of partitions that better reflect the overall structure.

A key challenge in this aggregation process lies in how to effectively measure the similarity—or structural affinity—between groups. A natural and commonly used factor is the relative edge density across groups. One classical metric in this regard is \textbf{RatioCut}, which evaluates the sum of edge weights between two groups normalized by their sizes. However, in signed networks, whether a set of groups forms a coherent and semantically consistent “community” also depends on the balance of the merged partition. As defined in Equation (2), a higher \textbf{BalanceRatio} indicates better preservation of signed structural semantics. To achieve this process, we employ an agglomerative hierarchical clustering strategy that progressively builds larger communities by aligning structural proximity and semantic consistency.
The specific formulas are as follows:

{\scriptsize
\begin{align}
\text{RatioCut}(C_i, C_j) = {} & \frac{\sum_{u \in C_i, v \in C_j} (A^+_{uv} + A^-_{uv})}{|C_i|} \nonumber \\
& + \frac{\sum_{u \in C_i, v \in C_j} (A^+_{uv} + A^-_{uv})}{|C_j|} \\
\text{BalanceRatio}(C_i \cup C_j) = {} & \frac{|\mathcal{T}_{\text{balanced}}|}{|\mathcal{T}_{\text{total}}|} \\
\text{Sim}(C_i, C_j) = {} & \alpha \cdot \text{RatioCut}(C_i, C_j) \nonumber \\
& + (1 - \alpha) \cdot \text{BalanceRatio}(C_i \cup C_j)
\end{align}
}

Here, $C_i$ and $C_j$ denote two groups (clusters), $\mathcal{T}_{\text{balanced}}$ and $\mathcal{T}_{\text{total}}$ represent the number of structurally balanced triads and the total number of triads, respectively, and $\alpha \in [0,1]$ is a trade-off coefficient.

Additionally, we control the edge balance of the resulting partitions. Unlike traditional partitioning methods that only consider balance in node count, we also account for balance in edge count.

\section{Experiments}

\begin{table*}[htbp]
  \centering \scriptsize
  \caption{Comparison of Macro F1 scores across different unlearning frameworks on four signed graph datasets and three SGNN backbones. \textbf{Bold} indicates the best-performing framework, and \underline{underline} indicates the runner-up. Scratch refers to retraining from scratch to serve as a performance upper bound.}
  \label{macro-f1-scores_0}
  \begin{tabular}{l|l|c|c|c|c|c|c}
    \toprule
    Dataset & Backbone & Scratch & GraphEraser & Random & GNNDelete & GIF & \textbf{SGU (Ours)} \\
    \midrule
    \multirow{3}{*}{Bitcoin-Alpha} 
      & SGCN  & 0.8291 ± 0.0051 & \underline{0.7755 ± 0.0078} & 0.7378 ± 0.0065 & 0.5266 ± 0.0024  & 0.6641 ± 0.0038  & \textbf{0.8059 ± 0.0083} \\
      & SNEA  & 0.8657 ± 0.0044 & \underline{0.8894 ± 0.0092} & 0.8871 ± 0.0087 & 0.5371 ± 0.0096  & 0.7121 ± 0.0088  & \textbf{0.9141 ± 0.0061} \\
      & SDGNN & 0.9260 ± 0.0038 & 0.8708 ± 0.0069 & \underline{0.8784 ± 0.0071} & 0.7067 ± 0.0073  & 0.7257 ± 0.0039  & \textbf{0.9239 ± 0.0097} \\
    \midrule
    \multirow{3}{*}{Bitcoin-OTC} 
      & SGCN  & 0.8517 ± 0.0063 & \underline{0.8233 ± 0.0057} & 0.8173 ± 0.0042 & 0.6444 ± 0.0038  & 0.7558 ± 0.0013  & \textbf{0.8548 ± 0.0094} \\
      & SNEA  & 0.9497 ± 0.0012 & 0.8046 ± 0.0086 & \underline{0.8239 ± 0.0074} & 0.7730 ± 0.0057  & 0.7944 ± 0.0068  & \textbf{0.8918 ± 0.0100} \\
      & SDGNN & 0.9706 ± 0.0059 & 0.8891 ± 0.0078 & \underline{0.9048 ± 0.0061} & 0.7971 ± 0.0098  & 0.7943 ± 0.0096  & \textbf{0.9282 ± 0.0043} \\
    \midrule
    \multirow{3}{*}{Epinions} 
      & SGCN  & 0.8775 ± 0.0029 & 0.7773 ± 0.0062 & \underline{0.7844 ± 0.0049} & 0.7009 ± 0.0013  & 0.7813 ± 0.0054  & \textbf{0.8195 ± 0.0053} \\
      & SNEA  & 0.9026 ± 0.0037 & 0.8222 ± 0.0071 & \underline{0.8490 ± 0.0052} & 0.7680 ± 0.0010  & 0.7819 ± 0.0051  & \textbf{0.8942 ± 0.0085} \\
      & SDGNN & 0.9608 ± 0.0060 & 0.8735 ± 0.0067 & \underline{0.9068 ± 0.0046} & 0.8519 ± 0.0034  & 0.8548 ± 0.0017  & \textbf{0.9338 ± 0.0079} \\
    \midrule
    \multirow{3}{*}{Slashdot} 
      & SGCN  & 0.7860 ± 0.0055 & 0.6979 ± 0.0098 & \underline{0.7020 ± 0.0044} & 0.4596 ± 0.0077  & 0.5140 ± 0.0037  & \textbf{0.7372 ± 0.0067} \\
      & SNEA  & 0.8486 ± 0.0039 & 0.8380 ± 0.0077 & \underline{0.8488 ± 0.0089} & 0.7352 ± 0.0069  & 0.7399 ± 0.0027  & \textbf{0.8816 ± 0.0054} \\
      & SDGNN & 0.8973 ± 0.0072 & 0.8550 ± 0.0061 & \underline{0.8748 ± 0.0053} & 0.7759 ± 0.0041  & 0.7821 ± 0.0044  & \textbf{0.9065 ± 0.0086} \\
    \bottomrule
  \end{tabular}
\end{table*}

\subsection{Experimental Setup}
\noindent\textbf{Datasets}
 Our method are evaluated on four signed graph datasets of various sizes, including Bitcoin-Alpha~\cite{kumar2016edge, kumar2018rev2}, Bitcoin-OTC~\cite{kumar2016edge, kumar2018rev2}, Epinions~\cite{leskovec2010signed}, and Slashdot~\cite{leskovec2010signed}, 
all of which contain real-world interactions and are widely used in signed graph research. 
Bitcoin-Alpha and Bitcoin-OTC are trust networks, Epinions is a social network based on user ratings, 
and Slashdot is a technology-related online community. 


\noindent\textbf{SGNN Backbones}
We equip our method with three SGNN backbones, SGCN~\cite{derr2018signed}, SNEA~\cite{li2020learning}, and SDGNN~\cite{huang2021sdgnn}, to evaluate its versatility.

\noindent\textbf{Baselines}
There is no previous work on signed graph unlearning based on our literature review. To demonstrate the effectiveness of our method, baselines include three graph unlearning algorithms for general graph, i.e., GraphEraser ~\cite{chen2022graph}, learning-based GNNDelete~\cite{cheng2023gnndelete}, and influence function GIF~\cite{wu2023gif}.
We also introduced Scratch, a scheme involving retraining from beginning to end. 
In addition, Random is a random-based graph partition for signed graph.

\noindent\textbf{Metrics}
The utility of the model is assessed using the Macro F1 score, which takes into account the imbalance between the positive and negative edges in the dataset. Unlearning capability is evaluated using the Attack AUC of membership inference~\cite{chen2021machine}.

\noindent\textbf{Experimental Settings}
Each dataset as a whole graph is randomly split into two disjoint parts, where 80\% of edges are used in training SGNN models, and 20\% of edges are used to evaluate the model utility. By default,
we set the number of shards $k$ for Bitcoin-Alpha, Bitcoin-OTC, Epinions, and Slashdot to 10, 10, 50, and 50, respectively.
Our codes and datasets are available at \url{https://anonymous.4open.science/r/Signed_Graph_Unlearning-5A1D}.

\noindent\textbf{Implementation}
SGU is implemented using Python 3.10 and PyTorch 2.5.0. Experiments were conducted on a machine with an NVIDIA RTX 3060 GPU and an Intel Core i5-12400F CPU. All the experiments regarding model utility are run 10 times, and we report the mean and standard deviation.

\subsection{Model Utility of SGU}
We have conducted a comparative analysis of SGU against the results of four graph unlearning baselines on three SGNN backbones and four datasets.

The experimental results presented in Table~\ref{macro-f1-scores_0} reveal the following findings: 
(1) SGU outperforms the four baselines in terms of model utility. For example, on the EPINIONS dataset, SGU shows an average improvement of 7.03\% compared to GraphEraser, and on the Bitcoin-Alpha dataset, it outperforms Random Partitioning by 5.82\%. 
(2) Under certain settings, SGU demonstrates model utility comparable to Scratch, proving its practicality in terms of model utility.

\subsection{Unlearning Efficiency of SGU}

 First, we measure the time cost of the SGU partitioning process. Table~\ref{sgu-runtime} reports the time cost of the SGU's partitioning phase, which consists of two stages: Stage (a), \textit{Signed Structure Extraction}, and Stage (b), \textit{Signed-Aware Balanced Clustering}. The results show that the FOCG-based signed structure extraction demonstrates good scalability.

\begin{table}[htbp]
  \centering
  \scriptsize
  \caption{Time cost (in seconds) of SGU’s partitioning phase, including Stage (a) Signed Structure Extraction and Stage (b) Signed-Aware Balanced Clustering.}
  \label{sgu-runtime}
  \begin{tabular}{l|cccc}
    \toprule
    Dataset &  Bitcoin-Alpha & Bitcoin-OTC & Epinions &Slashdot  \\
    \midrule
    Stage(a) & 1.11& 2.69& 63.22&85.94\\
    Stage(b) &0.89& 1.25 & 42.51&53.67 \\
    \bottomrule
  \end{tabular}
\end{table}

Second, we compare the average unlearning time of SGU with several baselines.
As shown in Table~\ref{tab:unlearning-time}, SGU significantly outperforms Scratch by achieving roughly an order of magnitude speedup. Although GraphEraser reduces time consumption compared to Scratch, its unbalanced partitioning strategy leads to noticeably higher runtimes than SGU.
    
\begin{table}[htbp]
  \centering
  \scriptsize
  \caption{Average unlearning time (in seconds) comparisons on four signed graph datasets.}
  \label{tab:unlearning-time}
  \begin{tabular}{l|cccc}
    \toprule
    Method & Bitcoin-Alpha & Bitcoin-OTC & Epinions & Slashdot \\
    \midrule
    Scratch      & 42.77  & 61.54   & 1214.95 & 1702.64 \\
    GraphEraser  & 13.06  & 22.92   & 452.56  & 587.25  \\
    Random       & 4.87   & 5.91    & 144.37  & 164.09  \\
    GNNDelete    & 1.33   & 2.12    & 8.19    & 4.54    \\
    GIF          & 0.31   & 0.38    & 0.69    & 0.95    \\
    SGU (Ours)   & 5.62   & 7.27    & 173.59  & 202.61  \\
    \bottomrule
  \end{tabular}
\end{table}

\subsection{Effectiveness of Signed-Aware Balanced Clustering}
To isolate and verify the contribution of our proposed clustering method to the overall model utility, we further compare SGU with a variant named SGU Random. In this variant, the Signed-Aware Balanced Clustering is replaced by a random aggregation strategy. The experimental results summarized in Table~\ref{macro-f1-scores_1}.  
SGU consistently outperforms SGU Random across all datasets and backbones.


\begin{table}[htbp]
  \centering
  \scriptsize
  \caption{Macro F1 scores of SGU and SGU Random across different signed graph datasets and SGNN backbones. \textbf{Bold} indicates better performance.}
  \label{macro-f1-scores_1}
  \begin{tabular}{l|l|c|c}
    \toprule
    Dataset & Model & SGU & SGU Random \\
    \midrule
    \multirow{3}{*}{Bitcoin-Alpha} 
      & SGCN  & \textbf{0.8059 ± 0.0083} & 0.7924 ± 0.0057 \\
      & SNEA  & \textbf{0.9141 ± 0.0061} & 0.7839 ± 0.0073 \\
      & SDGNN & \textbf{0.9239 ± 0.0097} & 0.8124 ± 0.0065 \\
    \midrule
    \multirow{3}{*}{Bitcoin-OTC} 
      & SGCN  & \textbf{0.8548 ± 0.0094} & 0.8392 ± 0.0056 \\
      & SNEA  & \textbf{0.8918 ± 0.0100} & 0.8653 ± 0.0081 \\
      & SDGNN & \textbf{0.9282 ± 0.0043} & 0.8751 ± 0.0064 \\
    \midrule
    \multirow{3}{*}{Epinions} 
      & SGCN  & \textbf{0.8320 ± 0.0045} & 0.8195 ± 0.0053 \\
      & SNEA  & \textbf{0.8942 ± 0.0085} & 0.8217 ± 0.0062 \\
      & SDGNN & \textbf{0.9338 ± 0.0079} & 0.8892 ± 0.0050 \\
    \midrule
    \multirow{3}{*}{Slashdot} 
      & SGCN  & \textbf{0.7372 ± 0.0067} & 0.7235 ± 0.0051 \\
      & SNEA  & \textbf{0.8816 ± 0.0054} & 0.7920 ± 0.0093 \\
      & SDGNN & \textbf{0.9065 ± 0.0086} & 0.8320 ± 0.0056 \\
    \bottomrule
  \end{tabular}
\end{table}

\subsection{Unlearning Capability of SGU}
We assess the unlearning capability of SGU through a member inference attack (MIA) approach tailored for GNNs. 
In our experiments, we randomly remove 0.5\% of the edges and then apply MIA to compare the results under two conditions: one using the original GNN backbone (denoted as $A$), and the other after applying SGU (denoted as $A_s$). As presented in Table~\ref{tab:mia_auc}, the AUC values under $A_s$ drop markedly and converge toward 0.5, which is comparable to random prediction. This highlights the reduced leakage of information.

\begin{table}[htbp]
  \centering
  \scriptsize
  \caption{AUC scores of Membership Inference Attacks (MIA) on original models ($A$) and SGU models ($A_s$).}
  \label{tab:mia_auc}
  \begin{tabular}{l|cc|cc|cc}
    \toprule
    \multirow{2}{*}{Dataset} & \multicolumn{2}{c|}{SGCN} & \multicolumn{2}{c|}{SNEA} & \multicolumn{2}{c}{SDGNN} \\
    \cline{2-7}
     & $A$ & $A_s$ & $A$ & $A_s$ & $A$ & $A_s$ \\
    \midrule
    Bitcoin-Alpha     & 69.1 & 51.4 & 70.5 & 51.0 & 70.8 & 51.2 \\
    Bitcoin-OTC       & 72.8 & 52.0 & 73.9 & 51.2 & 74.2 & 51.3 \\
    Epinions          & 68.5 & 51.2 & 69.1 & 50.8 & 67.9 & 50.6 \\
    Slashdot          & 60.3 & 50.3 & 61.2 & 50.1 & 61.0 & 50.2 \\
    \bottomrule
  \end{tabular}
\end{table}

\section{Conclusions}
In this paper, we propose SGU, the first unlearning framework for signed graph neural networks (SGNNs). In the partition module, SGU adopts a bottom-up signed graph reconstruction approach by leveraging a quadratic optimization to extract meaningful signed structures, which are subsequently clustered through a sign-aware hierarchical strategy that integrates structural proximity and edge polarity. 
We further introduce edge-scale balance to improve partition quality for SGNNs. Extensive experiments demonstrate that SGU maintains high model utility and offers solid unlearning guarantees across various signed graph settings. Future work can study our singed graph unlearning in large-scale realworld applications.

\renewcommand{\refname}{}
\section{References}
\vspace{-2.0em}
\bibliographystyle{IEEEbib}
\bibliography{Template}

\end{document}